\documentclass[apj]{emulateapj}

\usepackage{graphicx}

\begin{document}

\title{The unusual radio afterglow of the ultra-long gamma-ray burst GRB\,130925A}

\author{
Assaf Horesh\altaffilmark{1}, 
S. Bradley Cenko\altaffilmark{2,3}, 
Daniel A. Perley\altaffilmark{4,5}, 
S. R. Kulkarni\altaffilmark{4}, 
Gregg Hallinan\altaffilmark{4}, 
and Eric Bellm\altaffilmark{4}
}

\altaffiltext{1}{Benoziyo Center for Astrophysics, Weizmann Institute of Science,
76100 Rehovot, Israel}
\altaffiltext{2}{Astrophysics Science Division, NASA Goddard Space Flight Center,
Mail Code 661, Greenbelt, MD 20771, USA}
\altaffiltext{3}{Joint Space-Science Institute, University of Maryland, College Park,
MD 20742, USA}
\altaffiltext{4}{Cahill Center for Astrophysics, California Institute of Technology,
Pasadena, CA 91125, USA}
\altaffiltext{5}{Hubble Fellow}

\begin{abstract}

GRB\,130925A is one of the recent additions to the growing family of
ultra-long GRBs (T90$ \gtrsim 1000$\,s). While the X-ray emission of
ultra-long GRBs have been studied extensively in the past, no
comprehensive radio dataset has been obtained so
far. We report here the early discovery of an unusual radio afterglow
associated with the ultra-long GRB\,130925A. The radio
emission peaks at low-frequencies ($\sim 7$\,GHz) at early
times, only $2.2$\,days after the burst occurred. More notably, the radio
spectrum at frequencies above $10$\,GHz exhibits a rather steep
cut-off, compared to other long GRB radio afterglows. This cut-off can
be explained if the emitting electrons are either mono-energetic or
originate from a rather steep, $dN/dE \propto E^{-4}$, power-law energy
distribution. An alternative electron acceleration mechanism may be required to produce such an electron energy distribution
Furthermore, the radio spectrum
exhibits a secondary underlying and slowly varying component. This may hint that the radio
emission we observed is comprised of emission from both a reverse and
a forward shock. We discuss our results in comparison with previous
works that studied the unusual X-ray spectrum of this event and
discuss the implications of our findings on progenitor scenarios. 

\end{abstract}

\section{Introduction}
\label{sec:intro}

Gamma-ray bursts (GRBs) are the most energetic explosions known in the
Universe. These events exhibit prompt gamma-ray emission followed by
what is referred to as afterglow emission in a wide range of wavelengths
from X-rays to radio. Currently, GRBs are classified based on
the duration of their prompt Gamma-ray emission. Short (duration $\lesssim 2$\,s) and long (duration $\gtrsim 2$\,s) bursts are believed to be a result of different
progenitor systems (See
 reviews by Woosley \& Bloom 2006; Nakar 2007; Berger 2014).

The common wisdom suggests that long GRBs originate from the
core-collapse of
massive stars (e.g., MacFadyen \& Woosley 1999). Short GRBs, on the other hand,
presumably arise from coalescence of two neutron stars (e.g.,
Paczynski 1986; Eichler et al. 1989). In these scenarios,
the afterglow is explained by the fireball model (Piran 1999;
Sari et al. 1999) where the broadband emission originates from electrons in the circumstellar (or interstellar)
medium (CSM or ISM) which are
accelerated by a forward shock driven by the relativistic ejecta. While many studies (see above) focus on understanding their origin, GRBs also serve as natural laboratories to study
physical processes in extreme conditions, such as relativistic particle
acceleration. In turn, understanding these processes and the
conditions in which they occur can provide clues as to the true nature of GRBs. This paper is related only to GRBs with
long prompt emission and
thus we will discuss only this type of GRBs, hereafter. 

Recently, Levan et al. (2014) have suggested that another class of
high-energy transients may exist, with possibly a different progenitor
system. They pointed out
that a handful of GRBs exhibit very long prompt
gamma-ray emission ($\geq 1000$\,s). These events are usually followed
by a late-time X-ray afterglow which shows flaring activity as late as $10^{4}$\,s  after
the initial burst. Unlike other long
GRB afterglows, the flaring activity in these ultra-long GRBs has
especially high flux and longevity.
Levan et al. suggest that these ultra-long GRBs may be a result of
either a tidal disruption event (TDE) of a star by a black hole or a
core-collapse of an extremely extended star (see also Gendre et al. 2013).

On 2013, September 25, another ultra-long GRB, namely GRB\,130925A, was discovered
(Lien et al. 2013). Both the X-ray and radio afterglows of this event
show unique features. In this paper, we discuss the unusual radio emission
of GRB\,130925A and discuss the implications of its unique
properties. We first summarize the details known so far from recent
studies of this GRB in \S~\ref{sec:GRB}. In \S
~\ref{sec:obs} we describe our radio observations and the data
reduction. The data analysis and modeling is performed in \S
~\ref{sec:model}. We then discuss our findings in
\S~\ref{sec:discussion} and summarize in \S~\ref{sec:summary}.

\section{GRB\,130925A - Discovery and recent studies}
\label{sec:GRB}

GRB\,130925A was discovered by the Burst Alert Telescope (BAT;
Barthelmy et al. 2005) onboard the {\it Swift} satellite (Gehrels et
al. 2004). Golenetskii et al. (2013) reported that the prompt Gamma ray
emission, observed by the Konus-wind satellite, lasted for
$\sim 4500$\,s and had a fluence of $5.0 \pm 0.1 \times 10^{-4}$\,erg
cm$^{-2}$. Adopting a redshift of $z=0.347$, as measured by Vreeswijk et
al. (2013), results in isotropic energy of $E_{iso}\approx 1.5 \times
10^{53}$\,erg. Compared to other ultra-long GRBs (Levan et al. 2014),
GRB\,130925A is the nearest event discovered so far and the most
energetic (see Table~\ref{tab:grbs}). The prompt Gamma-ray emission was followed by a spectacular
X-ray emission with rapid flaring at the first $10^{4}$ seconds of the
event. Once the flaring ceased, the X-ray emission decayed as a
smooth power-law, typical of normal long GRBs at this stage. However,
additional uncharacteristic properties of the X-ray emission have been revealed. 
\begin{table}[!ht]

\small
\caption{GRB\,130925A in comparison to previously discovered ultra-long GRBs}
\smallskip
\begin{center}
\begin{tabular}{cccc}
\hline
\noalign{\smallskip}
Name & $z$ & $E_{iso}$   & $T_{90}$ \\
&   & [erg] & [sec] \\
\noalign{\smallskip}
\hline
\noalign{\smallskip}

GRB\,101225A & 0.85 & $1.2\times 10^{52}$ & $> 7000$ \\
GRB\,111209A & 0.67 & $5.2\times 10^{52}$ & $\sim 10,000$ \\
GRB\,101225A & 1.77 & $7\times 10^{52}$ &$\sim 6000$ \\
GRB\,130925A & 0.35 & $1.5\times 10^{53}$ &$\sim 4500$ \\

\noalign{\smallskip}
\hline
\smallskip
\end{tabular} 
\end{center}
{\small
Notes - The properties of the previously discovered GRBs are adopted
from Levan et al. (2014). }     
\label{tab:grbs}
\end{table}

Bellm et al. (2014) analysed the X-ray 
data obtained by the  {\it Swift}, {\it NuSTAR}, and the {\it
  Chandra} satellites. Surprisingly, they found that the X-ray spectrum in the energy range $1-20$\,keV, can not be modeled
with a single power-law, as in essentially most normal long GRB
afterglows to date (however, see Starling et al. 2012, that report a
thermal X-ray component in some GRBs associated with SNe) . Bellm et
al. obtained satisfactory fits to the observed X-ray spectrum with several models
including: 1. two
power-law components;  2. a power-law $+$ an absorption line at
$~6$\,keV; 3. a power-law and a blackbody component. However, the physical interpretation of the models was inconclusive.
Recently, Piro et al. (2014) analyzed additional X-ray data from the
{\it XMM} space observatory. According to their analysis, the X-ray emission is
dominated by a blackbody emission and that only a small contribution
of the X-ray
emission is due to non-thermal synchrotron emission from a traditional
afterglow (an afterglow from an external forward shock). 
A different explanation for the late-time X-ray emission was suggested
by Evans et al. (2014). They argue that the emission is a result of
reflection from dust which resides far away from the GRB. 
Both Piro et al. (2014) and Evans et al. (2014) find that the lack of
(or weak) 
X-ray emission from the external shock suggests  an extremely
low circumburst density ($n\leq 0.1\,{\rm cm}^{-3}$).

In the optical regime, Greiner et al. (2014) reported the detection of
a flare, $300 - 400$\,s after the prompt emission. At late-times, Tanvir et al. (2013) observed GRB\,130925A with
the Hubble Space Telescope (HST). They detected an
optical afterglow with an offset of 0$\farcs 12$ from the host galaxy
center. This offset may disfavor a TDE scenario for this
event. However, Tanvir et al. also note that the host  galaxy is
disrupted and that this may be a sign of a recent merger. In this case, it
is possible that a massive black hole can be offset from the optical
light center, thus still leaving a TDE as a plausible scenario.

\pagebreak
\section{VLA observations of GRB\,130925A}
\label{sec:obs}

We observed GRB\,130925A with the Karl G. Jansky Very Large Array (VLA) under a Director Discretionary Time (DDT)
program (14A-435; PI Horesh). Our observations consist of multi-epoch observations starting $2.2$\,days after the
burst\footnote{Unfortunately, due to the U.S. government shutdown, the VLA operations ceased at a critical time. Our second
  epoch was conducted on the last night before the shutdown, but was
  limited to only low frequencies. Further observations resumed much
  later on, 43 days after the discovery of the GRB.}. Each observation was
performed using a varying set of average frequencies: $3.4$\,GHz
(S-band), $6.1$\,GHz (C-band), $9$\,GHz (X-band), $14.75$\,GHz
(Ku-band), and $22$\,GHz (K-band). 
The various observing epochs (starting on 2013 Sep 27 UT; see
Table~\ref{tab:obs}) were performed with the VLA being either in the A or B configuration.

In all of our VLA observations, we used 3C48 as a flux calibrator and
J0240-2309 as a phase calibrator. The data were then reduced using
both AIPS  and CASA (McMullin et al. 2007) standard routines. To estimate the accuracy of our
flux calibration and to check for any calibration errors we compare
the measured flux of our phase calibrator at the various observing
epochs. The wide-band spectra 
of the phase calibrator at different times are consistent within the
following wide-band flux calibration errors: $6.3\%$, $1.3\%$, $2.5\%$, $5\%$,
and $5\%$ in the S-, C-, X-, Ku, and K-bands, respectively. Note that
these calibration errors were calculated when using the full bandwidth
in each band (2-8\,GHz bandwidth) and have been adjusted for sub-band
measurements. The full set of our measurements is presented in Table~\ref{tab:obs}.

\begin{table}[!ht]

\scriptsize
\caption{Summary of JVLA radio observations of GRB\,130925A}
\smallskip
\begin{center}
\begin{tabular}{cccc}
\hline
\noalign{\smallskip}
Time & Frequency   & Flux   & Flux r.m.s \\
days &  [GHz] & [$\mu$Jy] & [$\mu$Jy] \\
\noalign{\smallskip}
\hline
\noalign{\smallskip}

    2    & 4.8  & 237  & 17 \\
    2    & 7.4  & 298   & 12 \\
    2    & 9.5  & 293   & 14 \\
    2   & 13.5 & 146   & 20 \\
    2   & 16.0  & 89  & 22 \\
    2   & 22.0  & 104  & 13 \\
    9    & 4.8  & 216 & 14 \\
    9    & 7.4  & 214  &   9 \\
    9    & 8.5  & 180  & 10 \\
    9    & 9.5  & 203  &  10 \\
   43   &  3.0  & 133 &  27 \\
   43    & 4.8  & 105 & 13 \\
   43    & 7.4   & 87  &  10 \\
   43   & 13.5  & 60  &  9 \\
   43   & 16.0   & 62 &   9 \\
   43   & 22.0  & 83  &  8 \\
   58   &  3.0   & 99  & 27 \\
   58   &  4.8   & 77  & 15 \\
   58   & 7.4   & 68  & 15 \\
   58   & 8.5   & 65  & 14 \\
   58   & 9.5   & 56  & 15 \\
   58   & 14.75  & 46 &  11 \\
   74   &  3.0   & 83  &  6 \\
   74   &  6.1   & 71  &  6 \\
   74   & 9.0   & 56  &  5 \\
   74   &14.75   & 42  &  8 \\
   94   & 3.4   & 59  & 13 \\
   94   & 6.1   & 47  &  7 \\
   94   & 9.0   & 34   & 7 \\
   94  & 14.75  &  33 &   7 \\
   94  & 22.0  &  33   & 8 \\
  108  &  3.5  & 39  & 12 \\
  108  &  6.1  &  50  &  6 \\
  108   & 9.0   & 41  &  6 \\
  108  & 14.75 &  44  &  6 \\
  108  & 22.0   & 38  & 10 \\
  499 & 6.1 & $< 24$ & -- \\
  499 & 9.0 & $< 24.3$ & -- \\

\noalign{\smallskip}
\hline
\smallskip
\end{tabular} 
\end{center}
{\small
Notes - Time is  given in days in days since the burst. The errors presented in the table represent the rms error from
each image. When given, limits are $3\sigma$ detection limits. Systematic flux calibration errors were added according to
\S 2. Both the rms error and the calibration should be combined in
quadrature.}     
\label{tab:obs}
\end{table}

\section{Radio Spectrum Analysis}
\label{sec:model}

\begin{figure}
\centering
\includegraphics[width=0.45\textwidth]{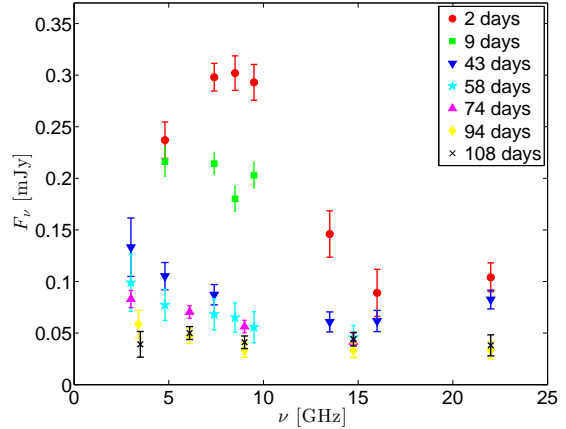} 
\caption{Radio spectra of GRB\,130925A at various VLA observation
  epochs. The initial spectrum shows a peculiar cut off at frequency
  $\geq 10$\,GHz. Moreover, the spectral evolution suggest the
  existence of an underlying (and slowly variable) constant-flux component. However, more
  than a year after its discovery, the radio emission from
  GRB\,130925A faded away below our detection limit (See Table~\ref{tab:obs}).}
\label{fig:radio_spec_all}
\end{figure}

Figure~\ref{fig:radio_spec_all} shows the observed wide-band radio spectra at
different epochs. The spectrum from the first epoch, $2.2$\, days after
discovery, shows that the emission is already peaking at
low frequencies ($\sim 7$\,GHz) and that it is 
strikingly cut off at $\geq 10$\,GHz. Both these properties are
not typical of normal GRB radio afterglows (see Chandra \& Frail
2012 for a review of GRB radio afterglow properties), in particular the
high-frequency cut off. 

Before discussing the implication of the high-frequency emission cut
off, we test whether this cut off can be due to some
modulation of the intrinsic flux via extreme interstellar scattering and scintillation (ISS). According to 
Cordes \& Lazio (2002), the Galactic scattering measure (SM)  towards the position of
the GRB is ${\rm log}\,SM\approx -3.69~[{\rm kpc~m}^{-20/3}]$. The
transition frequency from strong to weak scintillation is at $\approx
8$\,GHz. At higher frequencies only small ISS flux modulations are
expected. Thus it is unlikely that the observed cut off in the radio emission above
$10$\,GHz, is due to temporal strong scintillation (i.e., expected
modulation $\lesssim 10\%$). Moreover, even at frequencies below the transition frequency (i.e., the strong scattering regime), we do not observe strong scintillation (see \S~\ref{sec:scnit_radius}). 

In the context of the fireball afterglow model (see \S~\ref{sec:intro}),
GRB afterglow spectra are usually well described by a broken power-law
(Sari et al. 1998). 
This is a result of the radio emitting electrons being accelerated
into a power-law energy distribution, $dN/dE \propto E^{-p}$. The spectral behavior depends on some characteristic
frequencies, which define the transition between the different
spectral slopes (see a detailed description by Sari et al. 1998). In
short, at lower frequencies, the emission will be
optically thick. At frequencies where the emission becomes optically
thin, the specific flux will rise as $\nu^{1/3}$ up to some maximum
value after which the specific flux will decay as
$\nu^{-(p-1)/2}$. At even higher frequencies, beyond some cooling
frequency (which usually occurs at or above the optical regime, at early times), the specific
flux has a steeper decline of $\nu^{-p/2}$.

The typical average observed value of the electron energy power-law
distribution is $p\approx 2$ (e.g., Panaitescu \& Kumar 2001), but
with a relatively wide distribution of $\sigma_{p}\approx 0.5$ (Shen,
Kumar \& Robinson 2006). Thus, typically, the spectral slope of
the optically thin radio emission, above the peak frequency but below
the cooling frequency, is rather shallow ($\nu^{-0.5}$). The sharp
spectral cut off we observe in the case of GRB\,130925A,
raises the possibility that the power-law energy distribution of the
emitting electrons is much steeper than in any other observed GRB to
date. Another possibility is that an alternative model is required,
such as a mono-energetic energy distribution. We next test both models
by performing a minimum $\chi^{2}$ fitting to the data.

\subsection{The power-law energy distribution model}
\label{sec:power-law}

Here we  fit the initial observed radio spectrum with the
common power-law afterglow model, consisting of an optically thick synchrotron
self-absorbed emission ($\nu^{5/2}$) 
and an optically thin emission which has a power-law spectral
shape ($\nu^{-\beta}$).  Since, as seen in
Figure~\ref{fig:radio_spec_all}, the radio emission
appears to be decaying into a flat spectrum, it is possible that there is an additional
slowly varying flat spectrum emitting component which we treat as
constant over the time scale of our observations. Therefore, we perform the model fitting both with
and without a second constant flux component.
\begin{figure}
\centering
\includegraphics[width=0.45\textwidth]{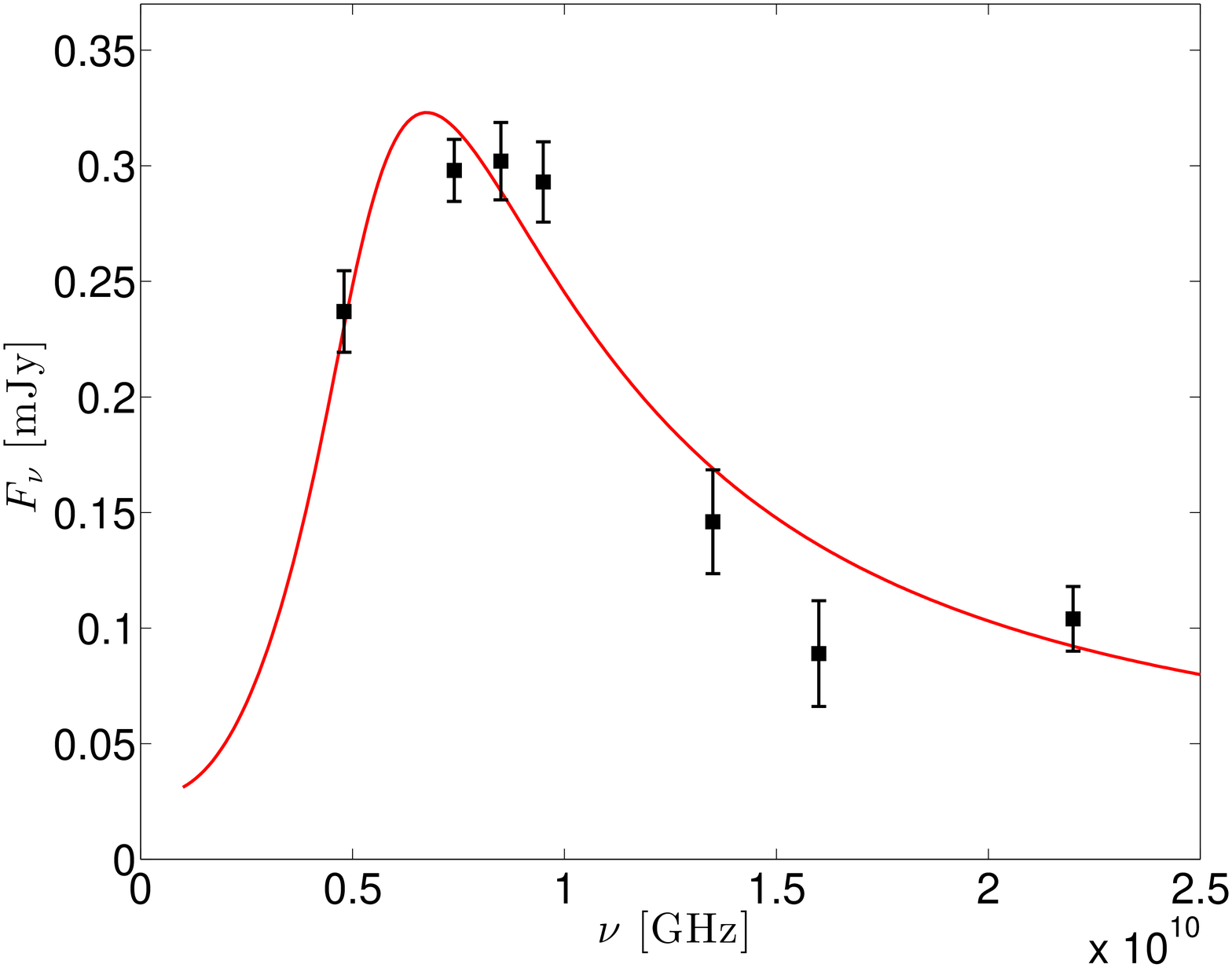}
\includegraphics[width=0.45\textwidth]{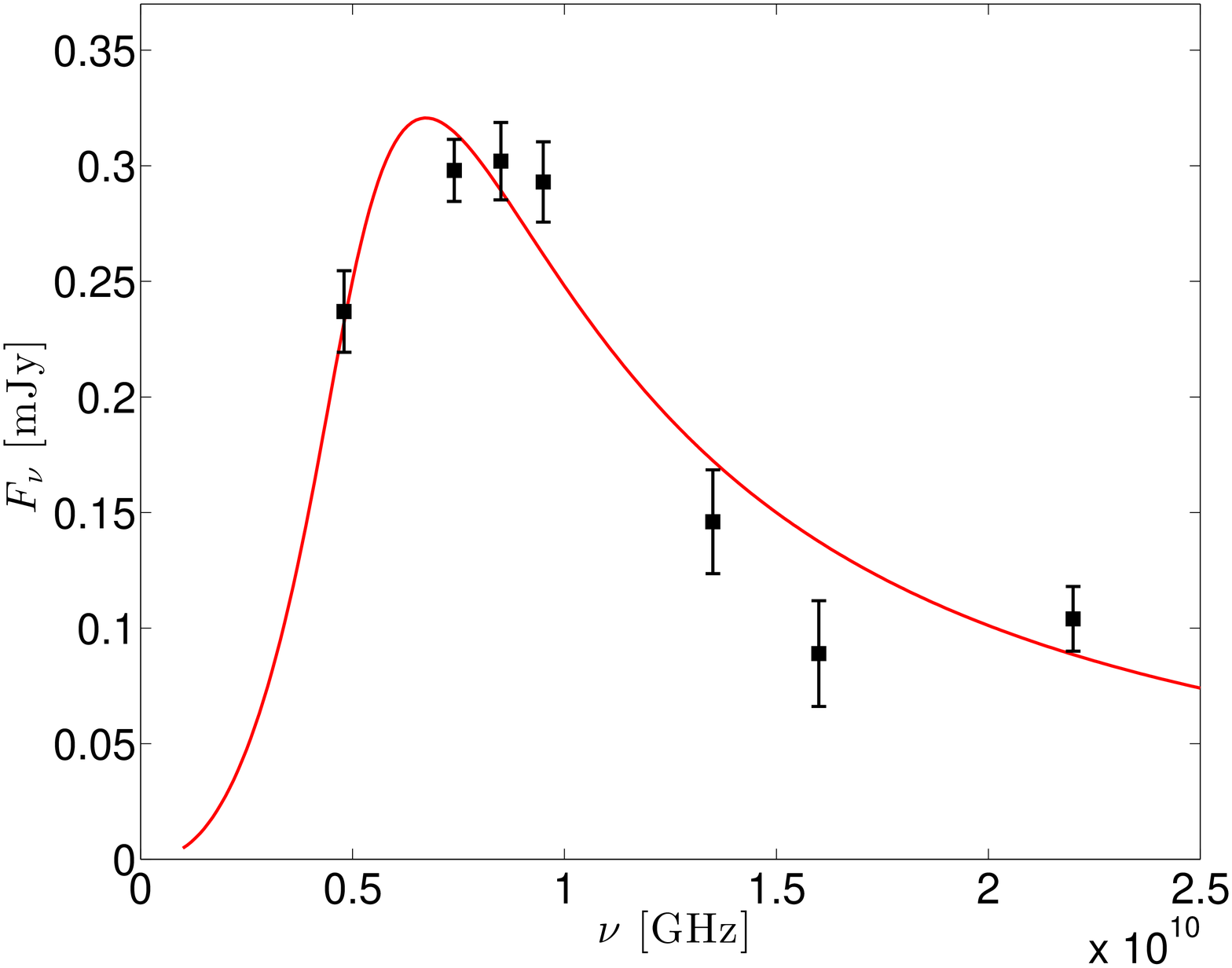}
\caption{Best fit of the power-law model to the observed initial
  radio spectrum. The two fits are with (left panel) and without
  (right panel) an
  additional constant flux component ($27\mu$Jy). Both fits suggest a steep
  electron energy distribution, not previously observed in GRBs (see
  text for details).}
\label{fig:power_law}
\end{figure}

Figure~\ref{fig:power_law} shows the best fit results. The
fit without a constant flux component has a reduced $\chi^{2}$, i.e. 
$\chi^{2}$ per degrees of freedom (d.o.f), 
of  $\chi^{2}_{\nu} = 4.2$ (d.o.f $= 3$). Adding a second component of constant flux to the fit results
in a  $\chi^{2}_{\nu}=6.2$ (d.o.f $=2$). The best fit spectral indexes in the former
and latter fits are $s= 1.4\pm 0.1 $ and $s= 1.6\pm 0.2$,
respectively. Assuming\footnote{The assumption here is that the
  cooling frequency is above the radio bands, an assumption which is
  supported by the data (see \S~\ref{sec:energy_dis})}  that $s=(p-1)/2$, suggests that the electron energy distribution
power-law indexes are $p\approx 3.8$ and $p\approx 4.2$, respectively. These are rather
steep energy distributions, not observed so far in GRB radio afterglows.

\subsection{A mono-energetic synchrotron emission}
\label{sec:mono}

Motivated by the sharp cut off observed in the initial radio spectrum,
we next consider an alternative model in which the observed synchrotron emission
originates from relativistic electrons with a mono-energetic energy
distribution. Since this model is rarely used for GRB afterglows
(however see Waxman $\&$ Loeb 1999), we next describe it in more
detail.

The synchrotron emission from a single electron is 
\begin{equation}
P_{\nu}(\nu)=\frac{\sqrt{3} q_{e}^{3} B}{m_{e} c^{2}}
F \left(\frac{\nu}{\nu_{c}}\right)
\end{equation}
 in the shockwave frame, where $c$ is the speed of light, $B$ is
 the magnetic field strength, $q_{e}$ and $m_{e}$ are the electron
 charge and mass, respectively, and 
\begin{equation}
F(x)=x\int_x^\infty K_{5/3}\left(\zeta \right)d\zeta , 
\end{equation} where $K$ is the modified Bessel function. The synchrotron frequency, $\nu_{synch}$ is defined as 
\begin{equation}
\nu_{synch}=\frac{3 \gamma_{e}^{2} q_{e} B}{4\pi m_{e} c}, 
\end{equation} 
where $\gamma_{e}$ is the Lorenz factor of the electrons (in the
shockwave frame).
In the relativistic case, the emission measured by an observer will be beamed and therefore in
the observer frame the emission is
\begin{equation}
P_{\nu,{\rm beamed}}(\nu)=\Gamma\frac{\sqrt{3} q_{e}^{3} B}{m_{e} c^{2}}
F\left(\frac{\nu}{\nu_{synch}}\right), 
\end{equation}
where $\Gamma$ is the bulk Lorentz factor of the shockwave. 
The frequency $\nu_{synch}$  in the observer frame will be
the same as above multiplied by a factor of $\Gamma$. 

In the case where internal synchrotron self-absorption (SSA) is dominant,
the observed specific luminosity in the shockwave frame will be
\begin{equation}
L_{SSA}(\nu)=\frac{4\pi R^{3}}{\eta} P_{\nu}
\left(\frac{1-e^{-\tau_{SSA}}}{\tau_{SSA}}\right).
\end{equation} 
Here, we assumed a planar absorption and the optical
depth is defined as 
\begin{equation}
\tau_{SSA}=\alpha_{SSA} \frac{R}{\eta},
\end{equation} where $\alpha_{SSA}$ is the absorption coefficient
defined as (Rybicki \& Lightman 1986):
\begin{equation}
\alpha_{SSA}=-\frac{c}{8\pi m_{e} \nu^{2}} \int \frac{N(E)}{E^2}
\frac{d}{dE} \left[E^{2} P_{\nu} \right]dE, 
\end{equation} where $N(E)$ is the volumetric density of electrons with
energy $E=\gamma_{e} m_{e} c^{2}$.

The shape of a SSA spectrum therefore depends also on the energy
distribution of the electrons, $N(E)$. In the mono-energetic case, $N(E)=N_{0}\delta(E-E_{0})$ (also assuming constant spatial
density). In this case, equation $7$ is reduced to: 
\begin{equation}
\alpha_{SSA}=-\frac{c^{2}}{8\pi \nu^{2}} \left[ N_{0}
  \left. \frac{dP_{\nu}}{dE}\right|_{E_{0}} + \frac{2
  P_{\nu}(E_{0})N_{0}}{E_{0}} \right].
\end{equation}
The specific flux in this case can therefore be easily calculated
using the following properties: $B$, $\gamma_{e}$, $n_{e}$, and $R$.

Finally, the observed flux can be reduced to the simple form (see Waxman \&
Loeb 1999): 
\begin{equation}
f_{\nu}=A\nu^{2}\zeta\left(\frac{\nu}{\nu_{synch}}\right),
\end{equation}
where 
\begin{equation}
\zeta\equiv \frac{1-e^{-\tau_{\nu}}}{1-[d\,ln\,F(x)/d\,ln\,x]}, 
\end{equation} 
\begin{equation}
A\equiv 4\pi \gamma_{e} m_{e} (R/D_{L})^{2}, 
\end{equation} and $D_{L}$ is the luminosity distance. 
 
\begin{figure}
\centering
\includegraphics[width=0.45\textwidth]{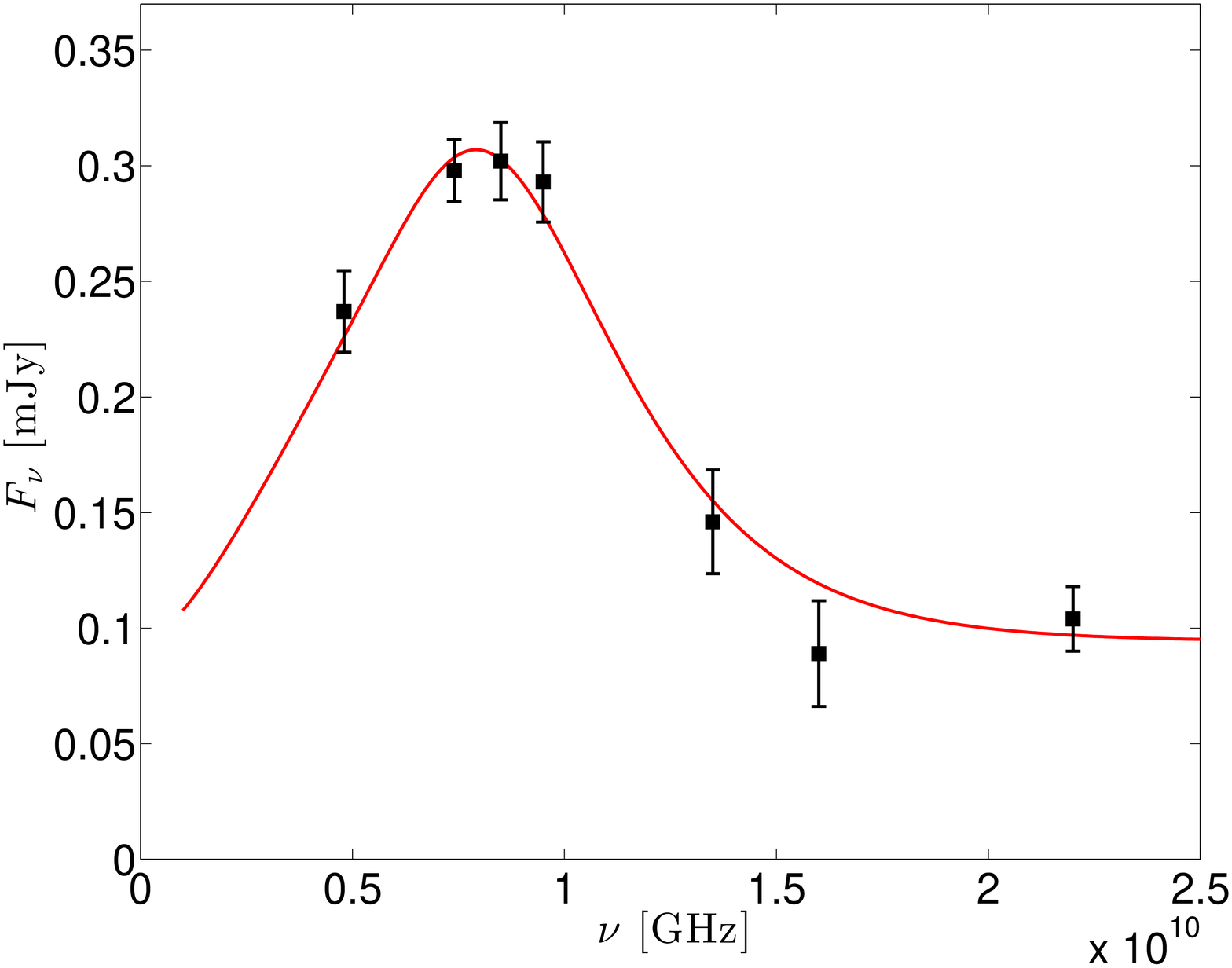}
\includegraphics[width=0.45\textwidth]{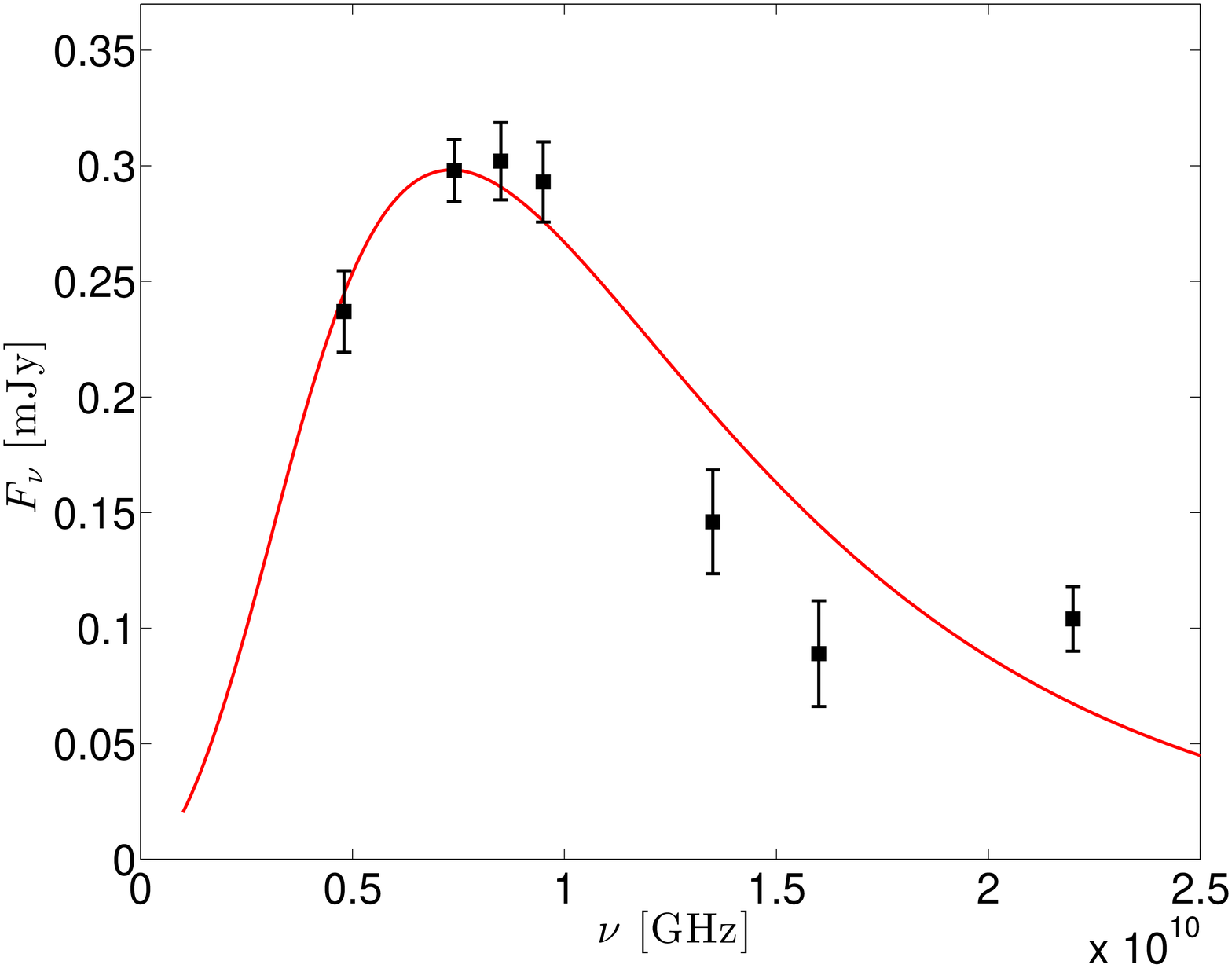}
\caption{Best fit of the mono-energetic model to the observed initial
  radio spectrum. The two fits are with (left panel) and without
  (right panel) an
  additional constant flux component ($95\mu$Jy). The mono-energetic + a constant
  flux component provide the best fit to the data. }
\label{fig:mono_fit}
\end{figure}

As in \S~\ref{sec:power-law}, we performed the model fitting twice, with and without an additional
constant flux component. The fitting of these two cases resulted in
$\chi^{2}_{\nu}=1.7$ (d.o.f $= 2$) and $\chi^{2}_{\nu}=6.2$ (d.o.f $= 3$), respectively, where the best fits are presented in
Figure~\ref{fig:mono_fit}. From a $\chi^{2}$ point of view,
the mono-energetic model with a secondary constant flux component is
slightly preferable (although not significantly) to the power-law model from
\S~\ref{sec:power-law}.

\subsection{Properties of the emitting medium}
\label{sec:prop}

We derive the electron density, the magnetic field strength and the
radius of the emitting regions for two cases: 1. An isotropic sub relativistic emitting sphere. 
2. Emission from a  relativistic jet.

\subsubsection{The sub relativistic isotropic case}

Here we assume that by the time of the observation the bulk motion is sub relativistic, and that the
emission is isotropic. In addition we also assume some ratio between the energy
density of the 
electrons ($E_{e}$) to the energy
density of the magnetic field ($E_{B}$).
Adopting the preferred model of mono-energetic
electrons and assuming equipartition ($E_{e}=E_{B}$), the best
fit to the data at the first observing epoch requires a radius of $R_{iso}\approx 8.5\times
10^{16}$\,cm, a 
large electron density of $n_{e}\approx 400$\,cm$^{-3}$ with
$\gamma_{e}\approx 45$, and a magnetic field strength of $B\approx
0.61$\, Gauss. The minimum required energy in this scenario is quite
large, $E_{min}\approx 2.3 \times 10^{49}$ erg. The fact that we observe only the optically thin synchrotron
emission from the third epoch and on does not allow us to obtain
estimates of the above model parameters in these additional epochs.

\subsubsection{A Relativistic Jet}

Recently, Barniol Duran, Nakar \& Piran (2013), presented a straightforward method
for calculating the above properties if the emission is originating
from a relativistic jet\footnote{The steepness of the electron energy distribution
  has only a small effect on the values of the derived properties
  (Barnio Duran; private communication)}. Adopting their prescriptions and assuming
equipartition we
derive the values of the emitting region properties using our measurements of the peak
flux and frequency at our initial epoch of observation. We find that, at that time, the jet is only mildly relativistic with a bulk Lorentz
factor of $\Gamma\approx 4$. The Jet radius is $R_{jet}\approx
1.3 \times 10^{17}$\, the electron density (assuming constant density)
$n_{e}\approx 11$\,cm$^{-3}$, the magnetic field strength is $B\approx
0.05$\,G, the electron Lorenz factor is $\gamma_{e}\approx 245$, and
the minimum kinetic energy is $E_{k}\approx 1.7\times 10^{47}{\rm
  erg}$. The relatively low value of $\Gamma$ should not be surprising
given that the initial Lorentz factor is estimated to be $\Gamma_{0}
\sim 20 - 37$ (Greiner et al. 2013). 

\subsubsection{A constraint on source size via scintillation}
\label{sec:scnit_radius}

As was already discussed above, radio flux variations, especially at
low frequencies, can be observed due to ISS. The ISS flux modulations
depends on the angular size of the emitting source. In turn, the
detection or lack of such modulations can be used to constrain the
source size. In our initial observing epoch, we observed the source at
4.8\,GHz twice within $\sim 2$ hours. We did not detect any strong
variation in the source flux between these two observations. Thus we
assume that the ISS modulations are within our measurement
errors. Conservatively, we estimate then that the ISS modulation is $\leq
20\%$. Using the equations from Walker (1998) for refractive and
diffractive scintillation, we find that the source angular size should
be $\geq 3\times 10^{17}$\,cm (radius $\geq 1.5 \times
10^{17}$\,cm), comparable to the source size we derived above.

\section{Discussion}
\label{sec:discussion}

\subsection{Electron energy distribution}
\label{sec:energy_dis}

In \S~\ref{sec:model} we found that the energy distribution of the radio emitting electrons can be fitted with
either a very steep power-law or a mono-energetic energy distribution. None of
these distributions have been observed in GRB
radio afterglows to date. Previously observed GRB afterglows have 
exhibited a power-law energy distribution with a power-law index of
$p\approx 2$. Theoretical studies (see Piran 2004 and references
therein) suggest that electrons are
accelerated at the GRB shock front via the Fermi process into a
power-law distribution with a power-law index of $p\approx 2.2-2.3$, in
agreement with past observations. A steep spectrum is possible if
the synchrotron cooling frequency is below the observed
frequency. However, the best fit of each of the two models in \S~\ref{sec:prop}
results in a cooling frequency of $\nu_{c}> 10^{12}$\,Hz, well above the observed
radio frequencies. Thus producing the observed steep energy
distribution may require an alternative particle acceleration
mechanism. Another recent challenge for current particle acceleration
models in GRBs has been made by Wiersema et al. (2014) who detected
circular polarization in the optical afterglow of GRB\,121024A.

Mono-energetic electrons have been observed in other astrophysical
sources, such as the galactic center (Lesch \& Reich 1992; Duschl \& Lesch 1994). In solar flares, the initial particle acceleration is believed
to be a result of magnetic reconnection which creates a mono-energetic
soft X-rays emission (see Benz 2008 for a review). Moreover, some
solar flares also exhibit steep optically thin radio spectra
(e.g. Nita, Gary, \& Lee 2004). 
Magnetic reconnection in accretion disks has also been suggested, as a possible
explanations for intra-day variability in AGNs and Blazars (e.g., Lesch
\& Phol 1992; Crusius-Waetzel \& Lesch 1998). 
Can
magnetic reconnection be involved in the case of GRB\,130925A? While
this question remains open, certain aspects of it are already being addressed by ongoing studies (e.g., Sironi \&
Spitkovsky 2014)

\subsection{The origin of the radio emission and its connection to the
X-rays}
\label{sec:origin_dis}

The radio emission from most long GRBs is believed to originate from a
forward shock in an external medium. If the radio emission of GRB\,130925A is indeed
originating from a forward shock then the properties found in \S~\ref{sec:prop}
are the properties of the ISM (or CSM). 
We test
whether the observed X-ray emission can arise from a simple forward
shock afterglow model, using the best fit parameters we derived from the
radio data. In this case, extrapolating the steep power-law (or mono
energetic) synchrotron emission into the X-ray bandpass results
in a low flux, orders
of magnitudes below the observed emission reported by Bellm et
al. (2013), Piro et al. (2014), and Evans et al. (2014) (expected $\nu F_{\nu}$
of $\sim 10^{-20}$\,erg\,cm$^{-2}$\,s$^{-1}$ compared to the observed
one of $\sim 10^{-12}$\,erg\,cm$^{-2}$\,s$^{-1}$). 

Both Piro et al. and Evans et al. suggest that there is only a weak
(or no) contribution from a forward shock to the observed X-ray
emission. Instead they provide different explanations for that
emission, such as blackbody radiation or scattering of the prompt
emission by dust. Both Piro et al. and Evans et al. conclude then that
the CSM density must be very low, with $n\leq 0.1\,{\rm cm}^{-3}$. In contrast, we find
that, if the radio emission originates from a forward shock, then the lack of non-thermal X-ray emission is not
due to low CSM density but rather due to the unusual steep energy
distribution of the emitting electrons. In fact the electron density we
find is higher by a factor of more than an order of magnitude than
that of Piro et al. (2014) and Evans et al. (2014).

Alternatively, it is possibile that the
radio emission is the result of a reverse shock ploughing through the
dense ejecta. In this scenario, the electron density we find is that
of the expanding ejecta shell. This renders the comparison between the
electron density we derive and the low density CSM environment found
by Evans et al. and Piro et al., irrelevant. In this scenario, the
underlying weaker component that is observed in the radio spectrum may
originate from the forwards shock. In fact, a second component of
$F_{\nu}\propto \nu^{1/3}$, instead of a constant flat spectrum one, is also
consistent with the data. Thus the existence of this second component
makes the reverse shock scenario more plausible. However, if this second component is not from forward shock emission, it means that the afterglow from the forward shock is suppressed. As was already discussed by Evans et al.,
one way to surpress forward shock emission is by invoking a very
weak magnetic field in the forward shock, compared to the one in the shocked
ejecta. Alternatively, this can also be achieved by reducing the
kinetic energy to less than $10^{50}{\rm erg}$, an energy budget which
is still in agreement with the minimum kinetic energy we find in
\S~\ref{sec:prop}. The reverse shock scenario, however, cannot
explain the fact the we still see emission, above the secondary
constant flux component, at $43$ days. Any emission from a reverse
shock is expected to decrease by at least $1 - 2$ orders of magnitude
compared to the emission observed at early times. Thus, our
observations at late-times hampers the reverse shock scenario.

The question, though, what is the origin of the X-ray emission, still
remains. If we put the implications of the radio data aside, a forward
shock as an explanation for the X-ray emission becomes unlikely also
when comparing the emission at a wavelength of $2.2\mu{\rm m}$ (Greiner et al. 2014) to
the X-ray flux, two days after discovery. The ratio between the two implies a relatively low
spectral index between $-0.3$ to $-0.6$, while we expect a steeper
spectrum due to the transition beyond the cooling frequency. In fact, the shockwave
parameters derived by Piro et al. suggest a $2.2\mu{\rm m}$ emission
much higher than the observed one. 

It is possible, in
some cases, to produce X-ray emission via the inverse-Compton (IC) process. For
example, a large enough reservoir of optical photons can be easily
up-scattered to X-rays by electrons with a Lorentz factor of $\gamma_{e}
\sim 50$. This possibility is especially intriguing since both the
radio and the X-ray spectrum exhibit similar steep spectral slopes. Using the measurements of Greiner et
al. as an estimate for the available optical and infrared photons, we
find that the expected IC X-ray emission is $\nu F_{\nu} \sim 5\times
10^{-14}\,{\rm erg}\,{\rm cm}^{-2}\,{\rm s}^{-1}$. The observed X-ray
emission, therefore, can not be accounted for by IC, since it is higher by almost two orders of magnitude than the expected
IC emission. 

Another possibility is that the X-ray emission originates
from a different process and region in the progenitor system, than the
radio emission. As
already mentioned, Piro et al. suggest that most of the X-ray emission is
blackbody emission originating from a compact radius of $10^{11}$\,cm. This
radius is much smaller than that of the radio emitting region. In
contrast, in the Evans et al. model the X-ray echoing dust lies far
away at pc scales. At the same time, one can argue that the fact that
the emission at both radio and X-ray exhibit an unusual steep spectrum,
may suggest a common origin. Additional analysis of the observed X-ray emission is beyond the scope of this work.

\subsection{Implications for the nature of the progenitor}
\label{sec:progenitor_dis}

As GRB\,130925A belongs to the ultra-long GRB subclass, we consider
the two main progenitor scenarios suggested for this class, namely a
TDE or an extreme collapsar. One of the best examples of a
relativistic TDE is {\it Swift}J\,1644+57 (Levan et al. 2011; Bloom et
al. 2011; Burrows et al. 2011). Comparing the radio properties of {\it
  Swift}J\,1644+57 (Zauderer et al. 2011; Berger et al. 2012) to
GRB\,130925A we find significant differences. First, {\it
  Swift}J\,1644+57 was more radio luminous, by a factor of $> 50$ than
GRB\,130925A. Second, the radio spectrum peak of  {\it
  Swift}J\,1644+57 traversed below $10$\,GHz only at very late times,
after $300$\,days. Furthermore, the external medium surrounding 
{\it Swift}J\,1644+57, is much denser. Overall, the radio properties
of {\it Swift}J\,1644+57 do not resemble in any way those of
GRB\,130925A. This, of course, is not proof that GRB\,130925A is not a
TDE, as {\it Swift}J\,1644+57 is the first discovery of an onset
of a TDE, but its properties may not be representative of the the whole population of relativistic
TDEs. Beyond the comparison of the properties of GRB\,130925A to the one of
the {\it Swift}\,1644+57, the properties of the former do not meet the
expectations of theoretical TDE models. According to Giannios \&
Metzger (2011), the minimal rise time of the radio flux will be
$t_{rise}\approx 30 - 200$\,days, depending on the black hole
mass. In contrast, the peak radio emission of GRB\,130925A is already declining
between 2 to 9 days, after discovery.

The radio emission does not pose a clear challenge to
the collapsar scenario. The radio can be explained as either a
forward or a reverse shock in this scenario. However, Piro et al. have
suggested that the progenitor is a blue super giant (BSG) star with very little or no
mass-loss. If the radio is originating from a forward shock, then a
denser CSM than the one found by Piro et al. is required. Thus a
progenitor with such low mass-loss as the one suggested by Piro et
al. is not possible. On the other hand, the reverse shock case does not
present a contradiction to the BSG progenitor scenario.

In both of the above progenitor models, the spectral cut off
in the radio, has to be explained, especially since it was not
observed before in TDEs or normal long GRBs (collapsars). In
fact, the unique radio spectrum that may require an alternative
electron acceleration mechanism, may, in turn, point to a different
progenitor system that has not been considered so far. For example, in
\S~\ref{sec:energy_dis}, we speculated that magnetic reconnection may be the mechanism
responsible for the acceleration of the radio emitting
electrons. Singh et al. (2014) have studied
the role of magnetic reconnection in accretion disk systems from
micro-quasar scales to blazars. They found that fast reconnection
originating from the central core, can play a part in the main energy
output of these type of objects. At the same time, they found that
this core mechanism cannot explain the energy output of GRBs and that
the afterglow emission may arise from a more distant location, such as
the jet. However, there is a suggestion by Giannios (2013) that
magnetic reconnection may take place in jets as well. Can it be that
GRB\,130925A is in fact not related to either a collapsar or a TDE
event? Another piece of the puzzle is that the radio spectrum
slowly decays to a flat spectrum in $\sim 100$\,days. Can this be a
hint that this event is actually related to an AGN activity? Assuming
so, the minimum variability time of $\approx 1$\,s in the prompt
emission (Greiner et al. 2014), suggest a BH mass of $<
10^{5}\,{\rm M}_{\odot}$. This low mass disfavors the AGN
scenario. Moreover,  $\sim 1.5$\,years after the GRB discovery, the seemingly constant
flat radio emission component has disappeared.

\section{Summary}
\label{sec:summary}

We report here for the first time, the early radio observation and
detection of an ultra-long GRB, GRB\,130925A. The early radio spectrum
obtained $2.2$\,days after the burst, has some unusual
properties. First, the radio emission peaks at $\sim 7$\,GHz, already at early
times. Even more surprising is the sharp spectral cut off at
$>10$\,GHz. We find that the data can be fitted with a SSA emission
model in
which the emission originates from either mono-energetic electrons or an
electron population with an unusually steep power-law energy
distribution. This may require an alternative acceleration mechanism
other than the one usually used in relativistic shock models of 
GRBs.

Having unusual
properties in various wavelengths, GRB\,130925A  may be of
a completely different nature than other GRBs. There is no clear concise
scenario that describes the overall properties of this
particular GRB. Certainly the usual fireball scenario used to explain
long GRBs does not capture the whole picture as the different pieces
of the puzzle (radio, optical and X-rays) cannot necessarily be put
together in this picture.

In the overall scheme of ultra-long GRBs, it seems that our radio data
provides another evidence that differentiates this type of events from
other normal long GRBs. However, since this is the first detailed
early radio spectrum obtained for an
ultra-long GRB, it is not clear whether GRB\,130925A is representative
of the ultra-long GRB population as a whole. Still, GRB\,130925A raises
some interesting questions regarding the properties and the progenitor
nature of ultra-long GRBs. One example, is the fact that ultra-long
GRBs exhibit steep X-ray spectra. Margutti et al. (2014) explains
this by invoking dust echoes on parsec scales, similar to the
explanation of Evans et al. (2014) in the case of GRB\,130925A. Is it
accidental that both the X-ray and radio spectra are soft? Or maybe
the X-ray and radio emission are originally connected to particles
accelerated by the same mechanism? These are open questions that we
hope to find answers to in future panchromatic (radio to X-ray)
studies of ultra-long GRBs. 

\begin{acknowledgments}

We thank R. Barniol Duran, T. Piran, E. Nakar, R. Sari, K. Mooley for
useful discussions. We thank the VLA staff for promptly scheduling
the observation of this target of opportunity. We also acknowledge the
use of the Astronomical Matlab Packages by Ofek (2014). 
The National Radio Astronomy Observatory is a facility of the National Science Foundation operated under cooperative agreement by Associated Universities, Inc.
Research leading to these
results has received funding from the EU/FP7 via ERC grant 307260; ISF, Minerva,
and Weizmann-UK grants; as well as the “Quantum Universe”
I-Core Program of the Planning and Budgeting Committee
and the Israel Science Foundation. S.B.C. acknowledges funding from
NASA grant NNH13ZDA001N. Support for D.A.P was provided by NASA through an award issued by JPL/Caltech, and through Hubble Fellowship grant HST-HF-51296.01-A awarded by the Space Telescope Science Institute.

\end{acknowledgments}

{}

\end{document}